\documentclass[%
 aip,
 jcp,
 amsmath,amssymb,
 reprint,%
]{revtex4-2}

\usepackage{mathtools}
\usepackage{graphicx}
\usepackage{dcolumn}
\usepackage{bm}

\usepackage[utf8]{inputenc}
\usepackage[T1]{fontenc}
\usepackage{mathptmx}
\usepackage{etoolbox}
\usepackage{orcidlink}
\usepackage{physics}

\makeatletter
\def\@email#1#2{%
 \endgroup
 \patchcmd{\titleblock@produce}
  {\frontmatter@RRAPformat}
  {\frontmatter@RRAPformat{\produce@RRAP{*#1\href{mailto:#2}{#2}}}\frontmatter@RRAPformat}
  {}{}
}%

\preprint{AIP/123-QED}

\usepackage{graphicx}
\usepackage{dcolumn}
\usepackage{bm}
\usepackage{subfigure}
\usepackage{multirow}
\usepackage{soul}
\usepackage[normalem]{ulem}
\usepackage{xcolor}
\usepackage{kantlipsum}
\usepackage{textcomp}

\usepackage{booktabs}
\newcolumntype{C}{>{$}c<{$}}
\AtBeginDocument{
\heavyrulewidth=.08em
\lightrulewidth=.05em
\cmidrulewidth=.03em
\belowrulesep=.65ex
\belowbottomsep=0pt
\aboverulesep=.4ex
\abovetopsep=0pt
\cmidrulesep=\doublerulesep
\cmidrulekern=.5em
\defaultaddspace=.5em
}
\usepackage{siunitx}
\usepackage{multirow}

\definecolor{amber}{rgb}{1,0.49,0}

\newcommand{\editor}[2]{%
  \expandafter\newcommand\csname #1note\endcsname[1]{%
    \textcolor{#2}{(\textbf{#1:} ##1)}}%
  \expandafter\newcommand\csname #1\endcsname[1]{%
    \textcolor{#2}{##1}}%
  \expandafter\newcommand\csname #1cancel\endcsname[1]{%
    \textcolor{#2}{\sout{##1}}}%
  \expandafter\newcommand\csname #1change\endcsname[2]{%
    \textcolor{#2}{\sout{##1} ##2}}%
  \newenvironment{#1text}{\color{#2}}{\color{black}}
}

\editor{FG}{blue}
\definecolor{verde}{rgb}{0.,0.6,0}
\editor{PP}{verde}
\editor{resub}{red}

\usepackage{mathrsfs}
\usepackage{placeins}

\begin{document}

\title{Self-interaction and transport of solvated electrons in molten salts}

\author{Paolo Pegolo\,\orcidlink{0000-0003-1491-8229}}
\email{ppegolo@sissa.it}
\affiliation{SISSA---Scuola Internazionale Superiore di Studi Avanzati, 34136 Trieste, Italy}

\author{Stefano Baroni\,\orcidlink{0000-0002-3508-6663}}%
\email{baroni@sissa.it}
\affiliation{SISSA---Scuola Internazionale Superiore di Studi Avanzati, 34136 Trieste, Italy}
\affiliation{CNR---Istituto Officina dei Materiali, SISSA unit, 34136 Trieste}

\author{Federico Grasselli\,\orcidlink{0000-0003-4284-0094}}
\email{federico.grasselli@epfl.ch}
\affiliation{ COSMO---Laboratory of Computational Science and Modeling, IMX, \'Ecole Polytechnique F\'ed\'erale de Lausanne, 1015 Lausanne, Switzerland}

\date{\today}

\begin{abstract}
The dynamics of (few) electrons dissolved in an ionic fluid---as when a small amount of metal is added to a solution while upholding its electronic insulation---manifests interesting properties that can be ascribed to nontrivial topological features of particle transport (e.g., Thouless' pumps). In the adiabatic regime, the charge distribution and the dynamics of these dissolved electrons are uniquely determined by the nuclear configuration. Yet, their localization into effective potential wells and their diffusivity are dictated by how the self-interaction is modeled. In this article, we investigate the role of self-interaction in the description of localization and transport properties of dissolved electrons in non-stoichiometric molten salts. Although the account for the exact (Fock) exchange strongly localizes the dissolved electrons, decreasing their tunneling probability and diffusivity, we show that the dynamics of the ions and of the dissolved electrons are largely uncorrelated, irrespective of the degree to which the electron self-interaction is treated, and in accordance with topological arguments.
\end{abstract}

\maketitle

\section{Introduction}\label{sec:introduction}

Extremely diluted alkali-metal/alkali-halide solutions feature solvated electrons released by the excess metal atoms that tend to localize in bound states analogous to polarons in dielectric solids.\cite{bredig1955miscibility,bronstein1958electrical,selloni1987electron,selloni1987localization, fois1988bipolarons, selloni1989simulation, lindemann1983cyclotron, popp1972diffusion, chaikin1972excitonic} 
Solvated electrons are the simplest anions in nature. They often appear as reaction intermediates in diverse chemical processes such as, e.g., radiolysis, photolysis, and electrolysis of polar materials.\cite{schindewolf1968formation} Despite having been observed for more than two centuries, since solutions of potassium in gaseous ammonia were examined by Sir Humphry Davy,\cite{davy1807bakerian} their properties are far from being completely explained, with some important advancements in their full understanding having appeared relatively recently in the literature, aided by the increasing accuracy and affordability of electronic-structure and machine-learning methods.\cite{marsalek2012structure,buttersack2020photoelectron,lan2021simulating,lan2022temperature} 

Solvated electrons in molten metal-metal halide solutions have been experimentally investigated especially since the 1940s, when molten salts were employed in the context of nuclear technologies.\cite{bredig1955miscibility,bredig1955miscibility2,bredig1955miscibility5,johnson1958miscibility,dworkin1962miscibility,bredig1963mixtures} Far from the NonMetal-to-Metal (NM-M) transition, the dynamics of such electronic states is adiabatic;\cite{fois1989approach, pegolo2020oxidation} therefore, the distribution of the excess electrons at each moment is entirely determined by the instantaneous ionic configuration, and the electronic motion is due to the ionic one.

The adiabatic variation of the potential energy surface determined by the nuclear dynamics is a natural playground for Thouless' theory of charge pumping;\cite{thouless1983quantization, niu1984quantised} in particular, the theorem of charge quantization, together with a recently discovered gauge invariance of transport coefficients,\cite{marcolongo2016microscopic,grasselli2019topological,grasselli2021invariance} provide a theoretical foundation for describing the charge-transport properties of ionic conductors according to the topology of their electronic structure. Notably, topological arguments demonstrate that in non-stoichiometric systems, which feature dissolved electrons, nontrivial charge transport can occur, meaning that adiabatic transport of charge can take place even without a net ionic displacement.\cite{pegolo2020oxidation, pegolo2022topology} This happens in non-stoichiometric molten salts such as metal/metal-halide solutions, where the electrical (ionic) conductivity can be recast as the sum of a part due to ions and one due to solvated electrons alone, the two contributions being uncorrelated from one another,\cite{pegolo2020oxidation, pegolo2022topology} resulting in a much increased electrical conductivity even before the NM-M transition.\cite{bronstein1958electrical}
The whole machinery behind these concepts is rooted in the modern theory of polarization;\cite{king1993theory, resta1994macroscopic} the latter provides also a means to rigorously characterize the electronically insulating state by exploiting its defining feature, i.e. the absence of dc conductivity, in terms of the localization of the electronic wavefunction.\cite{resta1999electron}

In practical calculations, electronic localization---and, vice versa, electronic diffusion---is determined by how self-interaction is accounted for in the employed theoretical framework. It is well known that standard Density Functional Theory (DFT) is affected by self-interaction errors due to the interaction of each electron with the total one-body electron density, including its own density.\cite{zhang1998challenge} The spurious contribution is partially removed by the approximate eXchange and Correlation (XC) functional, but errors are still large, especially for local and semi-local XC functionals.\cite{zhang1998challenge, bao2018self}
The excess electrons in metal-metal halide solutions are effectively few-electron systems and, as such, are particularly affected by spurious self-interactions.\cite{bao2018self}
Since fluids are structurally disordered materials, the localization of a small number of electrons is often facilitated with respect to crystalline systems, so even calculations employing semi-local XC functional result in rather localized electronic states whose dynamics has been studied from Ab Initio Molecular Dynamics (AIMD) simulations based on standard DFT.\cite{selloni1987electron, selloni1987localization, fois1988bipolarons, selloni1989simulation, fois1989approach, pegolo2020oxidation}
Nonetheless, the inclusion of a fraction of EXact (Fock) eXchange (EXX) naturally leads to a stronger degree of electronic localization, since EXX partially removes the effect of self-interaction, and helps describing, e.g., polarons in solids\cite{sio2019ab} and the localization in cavities of excess electrons in fluids.\cite{lan2021simulating}

It is thus expected that the EXX would quantitatively alter the charge transport properties of an ionic system containing solvated electrons, as the latter tend to be more localized and their interaction with the ionic species becomes more pronounced. Focusing on the paradigmatic case of non-stoichiometric molten salts, we show how the electrical conductivity remains separately determined by a purely ionic contribution and one uniquely due to the excess electrons' dynamics, the cross contribution still being vanishingly small. At the same time, we demonstrate that the effect of EXX is observed primarily when examining average structural and electronic properties, rather than instantaneous quantities.

\begin{figure}[tb]
    \centering
    \includegraphics[width=\columnwidth]{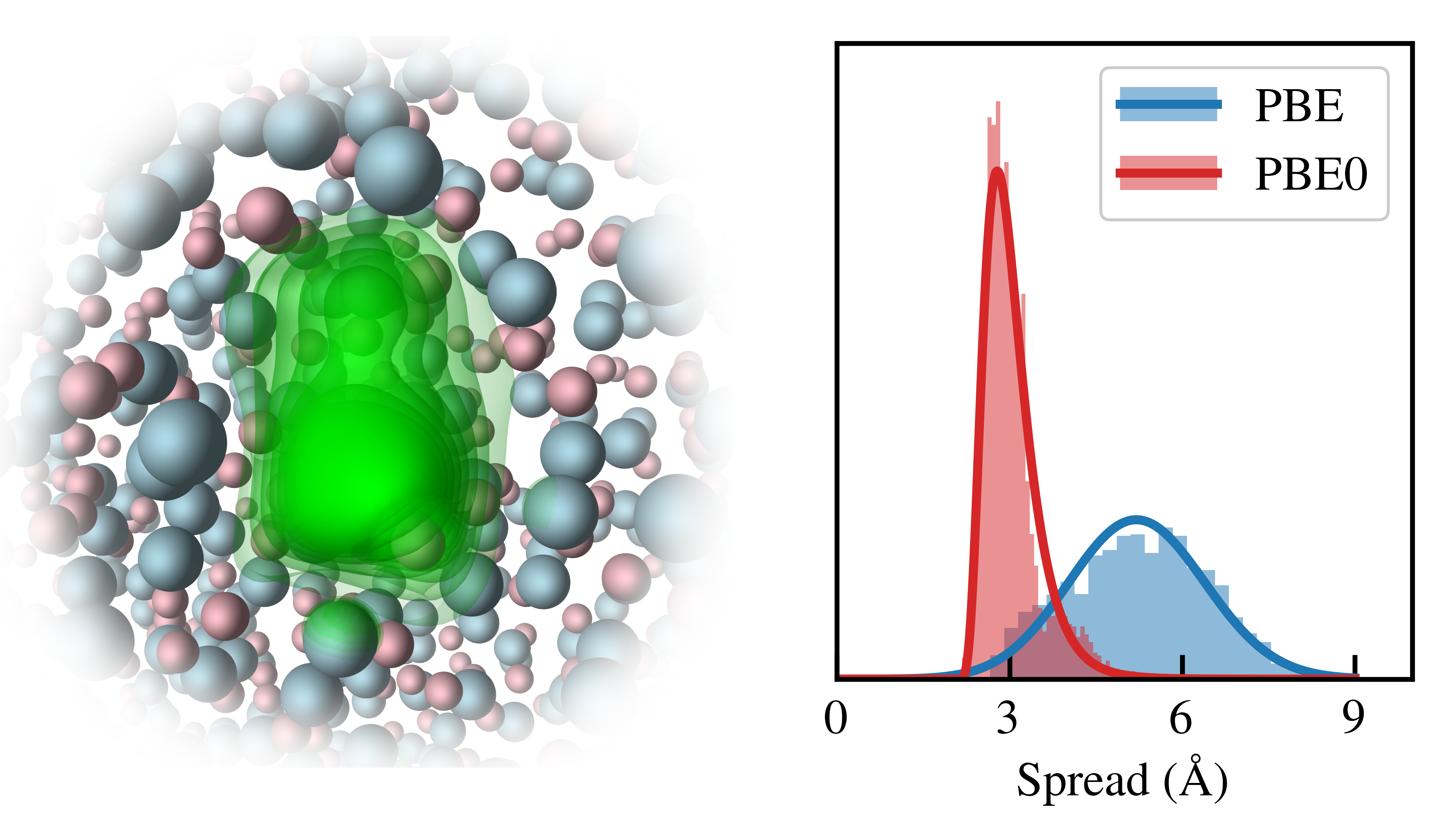}
    \caption{In the left panel, a typical snapshot taken from molecular dynamics simulations of non-stoichiometric molten NaCl. The volumetric data represents different isosurfaces of the bipolaron's charge density. Pink spheres represent Na nuclei, while light blue spheres Cl nuclei. In the right panel, the distribution (normalized histogram) of the bipolaron's spread computed with the PBE and the PBE0 functionals are reported in blue and red, respectively. The shaded bar-plots are the normalized histograms of the spread values; solid lines are log-normal fits to the histogram counts.}
    \label{fig:system}
\end{figure}

\section{Discussion}\label{sec:results}

We study the non-stoichiometric molten salt $\mathrm{Na}_{1+x}\mathrm{Cl}_{1-x}$, with $x \approx 0.06$, and we compare its properties using two different functionals. The first, chosen as our reference, is the semi-local PBE functional\cite{perdew1996generalized} that has been previously employed in \textit{ab initio} simulations of molten salts.\cite{pegolo2020oxidation} The second, hybrid, functional aims to enhance the localization of excess electrons. However, determining the appropriate amount of EXX to include in the calculation is often challenging. Advanced techniques have been developed to determine the optimal EXX fraction, able to provide an accurate description of excess electrons, especially polarons, in materials. These techniques involve approaches such as enforcing piecewise linearity on the DFT energy with respect to electron occupation, addressing in this way the self-interaction errors of DFT.\cite{falletta2022many, falletta2022polarons} Here, we have instead chosen to employ the widely-used PBE0 formulation,\cite{adamo1999toward} which is parameter-free and features 25\% of EXX. We have found this fraction of EXX to be sufficient to induce a suitable level of localization of the excess electrons that allow us to qualitatively compare the differences in properties of non-stoichiometric molten salts under the influence of EXX.

We focus on the simple case of 33 Na atoms and 31 Cl atoms. Simulations employing both functionals are carried out in a cubic cell with a side of $13\,\text{\r{A}}$ at a density of $1.40\,\mathrm{g/cm^3}$ and a temperature of $1300\,\mathrm{K}$. Further information is provided in Appendix~\ref{app:computational}. The presence of two extra Na atoms with respect to the stoichiometric formula leads to ionization and the release of one electron each, forming a solvated electronic pair called \emph{bipolaron}.\cite{selloni1987electron,selloni1987localization,fois1988bipolarons,fois1989approach,pegolo2020oxidation,pegolo2022topology,kristoffersen2018chemistry} The Highest Occupied Molecular Orbital (HOMO) corresponds to the bipolaron's wavefunction.\cite{pegolo2020oxidation,pegolo2022topology} To comprehend the bipolaron's characteristics, we determine the location and spatial extent of the HOMO. This can be achieved by transforming the Kohn-Sham Bloch states into a localized basis, such as the Wannier Functions (WFs). We can thus determine Wannier Centers (WCs) and Spreads (WSs), where the WS quantifies the spatial extent of the orbitals. In particular, the HOMO WC represents the bipolaron's position, while the HOMO WS provides insights into its degree of localization.\cite{resta1999electron, marzari2012maximally, pegolo2020oxidation} We employ the widely used Maximally Localized Wannier Functions (MLWFs)\cite{marzari2012maximally}, where the HOMO WS is defined as the variance of the position operator evaluated over the HOMO WF. In the following, we will loosely speak of spread referring also to its square root which, having the dimensions of a length, facilitates comparisons with distances. 

A typical snapshot taken from a simulation of non-stoichiometric molten NaCl is shown in the left panel of Fig.~\ref{fig:system}, where several isosurfaces of the HOMO charge density are colored in green. In the right panel, we show the distribution of the HOMO WSs computed during the two simulations.
Here, it is worth mentioning that the MLWF construction may underestimate the WS when the latter is not small compared to the simulation box size,\cite{li2023unambiguous} as it occurs when EXX is not considered. Therefore, the difference in the WS distribution between PBE and PBE0 may be even amplified, should the WS be estimated according to the accurate formulation of Ref.~\onlinecite{li2023unambiguous}. Including a portion of EXX in the XC functional impacts both structural and dynamical properties of non-stoichiometric NaCl. In the subsequent sections, we will address these aspects.

\subsection{Structural properties}\label{ssec:static}

\begin{figure*}[t]
    \centering   \includegraphics[width=1.9\columnwidth]{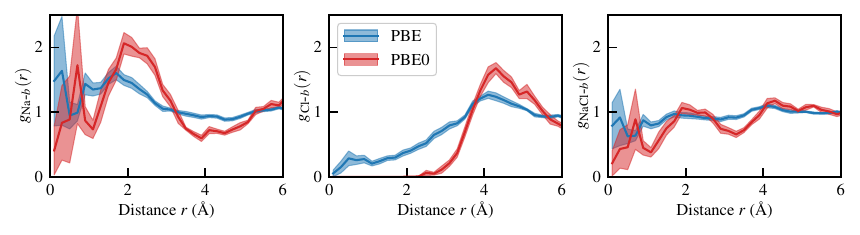}
    \caption{RPDFs of the bipolaron with Na (left), Cl (middle), and both (right), both with and without EXX. The shaded areas represent standard deviations obtained via a block average of $2.5\,\mathrm{ps}$-long segments of trajectory.}
    \label{fig:gofr}
\end{figure*}

To understand the structural effects of adding the EXX to the XC functional, we compute the Radial Pair Distribution Function (RPDF), $g(r)$, of the bipolaron (hereon labeled by $b$) and the atomic species in NaCl, that are shown in Fig.~\ref{fig:gofr}. We observe several distinct features that reveal the differences between PBE and PBE0. First, the likelihood of locating a Cl ion in close proximity to the bipolaron is significantly lower in the case of PBE0 as compared to PBE. Second, RPDF for the interaction between the bipolaron and any ion exhibits a much more pronounced structure with PBE0. This indicates the existence of a well-defined shell structure surrounding, on average, the bipolaron, which is notably absent when using the PBE functional. In contrast, PBE predicts a comparatively uniform distribution of ions around the bipolaron, lacking any discernible shell-like organization.
It must be noted that size effects can affect the results at large distances, since the RPDFs have not completely decayed within half the simulation cell's size. Features at shorter distances are nonetheless significantly different for PBE and PBE0.

\begin{figure}[tb]
    \centering
    \includegraphics[width=\columnwidth]{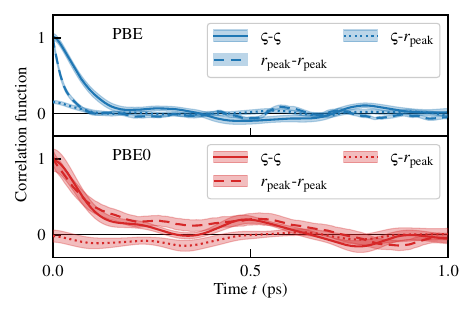}
    \caption{Time-correlation functions of the bipolaron's spread, $\varsigma$, and the first peak in the instantaneous bipolaron-sodium RPDF, $r_{\mathrm{peak}}$. The shaded areas represent the standard errors on the means computed via block-averaging over $1\,\mathrm{ps}$-long segments of trajectory.}
    \label{fig:correlation function spread}
\end{figure}

Since incorporating a fraction of EXX affects both the bipolaron's spread distribution (see the right panel of Fig.~\ref{fig:system}) and its local environment (see Fig.~\ref{fig:gofr}), we investigated the potential correlation between electronic and structural properties. 
To this end, we calculated the time-correlation functions of the (square root of the) bipolaron's spread, $\varsigma$, and the distance of the first peak in the instantaneous partial RPDF between the bipolaron and Na ions, $r_\mathrm{peak}$, which we use as a proxy for the bipolaron's local environment. 
The results are reported in Fig.~\ref{fig:correlation function spread}. The time-correlation functions were calculated using standardized time-series data for the spread and $r_\mathrm{peak}$, according to:
\begin{align}
    C_{AB}(t) = \frac{\left\langle (A(t)-\langle A \rangle) (B(0)-\langle B \rangle) \right\rangle}{\sqrt{\left\langle (A-\langle A \rangle)^2 \right\rangle \left\langle (B-\langle  B \rangle)^2 \right\rangle}},
\end{align}
with $A$ and $B$ being $\varsigma$ or $r_{\mathrm{peak}}$. The characteristic $\varsigma$--$\varsigma$ and $r_\mathrm{peak}$--$r_\mathrm{peak}$ correlation times appear to be the same for the PBE0 functional. For PBE, instead, the correlation time of $r_\mathrm{peak}$ is shorter than that of the spread, likely due to the fact that the erratic motion of the bipolaron in the presence of a semi-local functional make its local environment rapidly change in time. The cross-correlation function is relatively small in both cases, suggesting that the two quantities are nearly uncorrelated with each other.

This analysis suggests that the effect of EXX can be fully understood only by examining average quantities sampled throughout the entire dynamics, rather than focusing on instantaneous values. This is in accordance with the fact that the distribution of the values of the spread---a statistical quantity---is markedly different in the two simulations, but the allowed values---instantaneous quantities---are partially overlapping. As a result, by evaluating a single snapshot it is almost impossible to discern whether the simulation was performed using a PBE or PBE0 functional. This is confirmed by a Principal Component Analysis (PCA) performed on Smooth Overlap of Atomic Positions (SOAP) descriptors~\cite{bartok2013representing,de2016comparing,musil2021physics,musil2021efficient,goscinski2021role} associated with the bipolaron's local environments, whose results are presented in Appendix~\ref{app:data driven}.

\subsection{Transport properties}\label{ssec:dynamical}

Charge transport in electronically insulating fluids relies on ionic motion. Allowing ions to move enables charge displacement. In the linear regime, the electrical conductivity, $\sigma$, can be expressed by the Green-Kubo (GK) formula:\cite{green1952markoff,green1954markoff,kubo1957statistical,kubo1957statistical2,baroni2020heat}
\begin{align}\label{eq:gk sigma}
    \sigma = \frac{\Omega}{3 k_B T} \int_0^\infty \langle \mathbf{J}(t) \cdot \mathbf{J}(0) \rangle \dd{t},
\end{align}
or, equivalently, by the Helfand-Einstein formula:\cite{helfand1960transport, grasselli2021invariance}
\begin{align}\label{eq:helfand}
    \sigma = \frac{1}{3 \Omega k_\mathrm{B} T} \lim_{t \to \infty} \frac{1}{2t} \langle \abs{\bm{\Delta \mu}(t)}^2 \rangle.
\end{align}
Here, $\Omega$ represents the system's volume, $k_B$ is Boltzmann's constant, $T$ is the temperature, $\mathbf{J}$ denotes the charge flux, and $\bm{\Delta \mu}(t)=\Omega\int_0^t \mathbf{J}(t') \dd{t'}$ is the displaced dipole. For a quantum system in the adiabatic approximation, the definition of $\mathbf{J}$ must consider the quantum nature of electrons, while nuclei are treated as classical point charges. In principle, any partitioning of the continuous electronic charge density is equally valid. 
Within an independent-electron picture, $\mathbf{J}$ is usually defined in terms of either the macroscopic polarization\cite{resta1994macroscopic} and its derivatives with respect to nuclear coordinates, the Born Effective-Charge tensors,\cite{resta1994macroscopic, ghosez1998dynamical} or MLWFs.\cite{marzari2012maximally}

This perspective is quite different from the classical view of ionic fluids, where ions are considered point charges with a well-defined charge attached to them. This classical picture can be restored under suitable topological conditions by considering a combination of the gauge invariance of transport coefficients and Thouless' theory of quantization of particle transport. This approach allows the use of a charge flux defined in terms of integer atomic Oxidation States (OSs):\cite{grasselli2019topological, pegolo2022topology}
\begin{align}
    \mathbf{J}(t) = \frac{e}{\Omega} \sum_{\ell=1}^{N} Q_\ell \mathbf{V}_\ell(t).
\end{align}
Here, $Q_\ell$ and $\mathbf{V}_\ell$ represent the OS and velocity of the $\ell$th nucleus, respectively. This holds when the topology of the configuration space of nuclear coordinates does not contain relevant regions where the electronic gap closes, and the system becomes metallic; a class of systems where this holds is that of stoichiometric molten salts.\cite{grasselli2019topological, pegolo2022topology} Conversely, when this condition is not met, charge is displaced not only as OSs attached to nuclei, but also through adiabatic electronic diffusion, as seen in the case of solvated electrons in molten salts.\cite{pegolo2020oxidation, pegolo2022topology} A classical picture can still be maintained from the perspective of MLWFs: in the Wannier representation, the charge flux becomes:
\begin{align}\label{eq:J wannier}
    \mathbf{J}(t) = \frac{e}{\Omega} \left[\sum_{\ell=1}^{N} Z_{\ell} \mathbf{V}_{\ell}(t) - 2 \sum_{j=1}^{N_{\mathrm{el}}/2} \dot{\mathbf{R}}^{\mathrm{(W)}}_j \right],
\end{align}
where $\mathbf{R}_j^{(\mathrm{W})}$ refers to the position of the Wannier center associated with the $j$th occupied electronic band, and $Z_{\ell}$ is the nuclear (core) charge of the $\ell$th nucleus.\cite{marzari2012maximally, resta2021faraday, pegolo2022topology}

\begin{figure}[tb]
    \centering
    \includegraphics[width=\columnwidth]{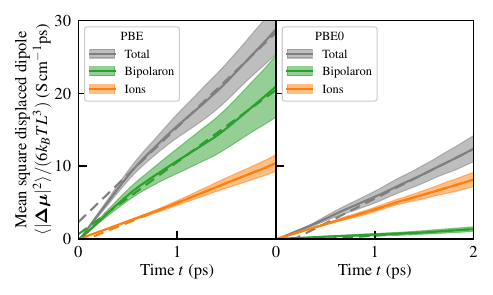}
    \caption{MSDD of non-stoichiometric molten NaCl computed with the PBE or the PBE0 functional. The shaded areas represent standard deviations on the mean computed via block averages.}
    \label{fig:sigma}
\end{figure}

It was shown in Refs.~\onlinecite{pegolo2020oxidation,pegolo2022topology} that, for the calculation of the electrical conductivity, we can employ the following (alternative) definition for the total charge flux:
\begin{align}\label{eq:flux oss and homo}
    \mathbf{J}(t) = \mathbf{J}_\mathrm{ions}(t) + \mathbf{J}_b(t),
\end{align}
where the flux assiociated to ions is $\mathbf{J}_\mathrm{ions}(t) \equiv \frac{e}{\Omega} \sum_{\ell=1}^{N} Q_{\ell} \mathbf{V}_{\ell}(t)$, and the one associated with the bipolaron is $\mathbf{J}_b(t) \equiv -2e\Omega^{-1}\dot{\mathbf{R}}_{\mathrm{HOMO}}^{\mathrm{(W)}}(t)$.
In essence, here the charge flux is expressed as integer atomic OSs for the nuclei in the system, with $+1$ for Na ions and $-1$ for Cl ions (fist term at RHS), supplemented by the neutralizing effect of a solvated bipolaron with an ``oxidation state'' of $-2$ and a velocity corresponding to the time-derivative of the HOMO WC position (second term at RHS).\cite{pegolo2020oxidation} The corresponding displaced charge dipoles can be obtained by time integration of the charge fluxes, and then substituted in Eq.~\eqref{eq:helfand}. The resulting total electrical conductivity is in general the sum:
\begin{equation}
    \sigma = \sigma_\mathrm{ions} + \sigma_{b} + \sigma_\mathrm{cross}
\end{equation}
where $\sigma_\mathrm{cross} \propto \int_0^\infty \langle  \mathbf{J}_\mathrm{ions}(t) \cdot \mathbf{J}_b(0) \rangle \dd{t}$.

The expectation values appearing here, denoted by angled brackets, are estimated by block-averaging over trajectory segments and, within each block, via an average over initial times. This is implemented in the software \texttt{analisi}.\cite{bertossa2022analisi}
Fig.~\ref{fig:sigma} displays the plot of the Mean Square Displaced Dipole (MSDD), ${(6 L^3 k_\mathrm{B} T)^{-1} \langle \abs{\bm{\Delta \mu}(t)}^2 \rangle}$, as a function of time for various displaced dipoles related to the ions alone and to the bipolaron.
The purely ionic contributions, whose slope are the respective values of $\sigma_{\mathrm{ions}}$, are comparable in both simulations. However, the bipolaron's contribution, $\sigma_{b}$, in the PBE0 case is significantly smaller than that in the PBE simulation, where it played a leading role in determining the electrical conductivity value. In fact, the bipolaron diffusivity is reduced from $38 \cdot 10^{-4}\,\mathrm{cm^2\,s^{-1}}$ for PBE to $3 \cdot 10^{-4}\,\mathrm{cm^2\,s^{-1}}$ for PBE0, while the ionic diffusivity is of the order of $10^{-4}\,\mathrm{cm^2\,s^{-1}}$ in both cases. Despite the reduced bipolaron contribution due to the inclusion of EXX, the total conductivity is consistent with the sum of the ionic and bipolaronic contributions, as already demonstrated for PBE in non-stoichiometric KCl in Ref.~\onlinecite{pegolo2020oxidation} and confirmed here, thus showing the lack of correlation between the two. The total ionic conductivity is significantly reduced after the inclusion of EXX, going from $\sigma \approx 18\,\mathrm{S\,cm^{-1}}$ in the PBE case to $\sigma \approx 5\,\mathrm{S\,cm^{-1}}$ in the PBE0 case. The latter compares fairly well with experimental conductivity measurements on Na-NaCl melts at similar metal concentration,\cite{bronstein1958electrical} while the former is appears rather overestimated.

To determine whether any form of dynamical correlation can exist, at least locally, we computed the cross-contribution to the conductivity between the bipolaron's charge flux and a \emph{local ionic flux}. The latter is defined as:
\begin{align}\label{eq:local flux}
    \mathbf{J}_{\mathrm{loc}}(\lambda, t) = \frac{e}{\Omega} \sum_{\ell=1}^N Q_\ell \mathbf{V}_\ell(t) \Theta(\lambda - |\mathbf{R}_\ell(t)-\mathbf{R}_{\mathrm{HOMO}}(t)|),
\end{align}
where $\Theta(x)$ is the Heaviside step-function, and $\lambda$ is some distance cutoff. Simply put, Eq.~\eqref{eq:local flux} contains the contribution to the charge flux due to ions within a distance $\lambda$ from the bipolaron, at each instant. A large value of $\lambda$ (up to half the simulation cell's side) entails computing the entire ionic flux, while a small value of $\lambda$ yields a quantity that depends only on the neighborhood of the bipolaron's position. The correlation between ionic and bipolaronic contributions to the electrical conductivity is estimated from the total local (i.e., $\lambda$-dependent) electrical conductivity, that we indicate with $\overline{\sigma}$ in order to distinguish it from the true electrical conductivity, $\sigma$. The local conductivity can be computed from Eq.~\eqref{eq:gk sigma} as
\begin{align}\label{eq:local sigma}
    \overline{\sigma}(\lambda) \propto \int_0^\infty \langle (\mathbf{J}_{\mathrm{loc}}(\lambda, t) + \mathbf{J}_b(t)) \cdot (\mathbf{J}_{\mathrm{loc}}(\lambda,0) + \mathbf{J}_b(0)) \rangle \dd{t}.
\end{align}

\begin{figure}[tb]
    \centering
    \includegraphics[width=\columnwidth]{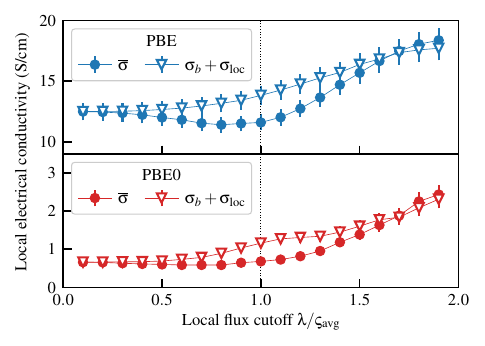}
    \caption{Local electrical conductivity of molten non-stoichiometric NaCl computed from the local charge flux of Eq.~\eqref{eq:local flux} as a function of the ratio between the local flux cutoff and the average bipolaron's spread. Filled circles indicate \emph{bona fide} GK results obtained from Eq.~\eqref{eq:local sigma}; empty triangles indicate the conductivity computed neglecting the correlation between the local ionic flux and the bipolaron's motion. Note that the $y$-axis of the upper panel starts from $9\,\mathrm{S\,cm^{-1}}$.}
    \label{fig:cross conductivity}
\end{figure}

Due to the relatively short PBE0 trajectories available, we employ the efficient cepstral analysis technique\cite{ercole2017accurate} as implemented in \textsc{SporTran}\cite{ercole2022sportran} to obtain the conductivity value from the fluxes' time-series.
Expanding the sums in the correlation function in Eq.~\eqref{eq:local sigma} enables us to separate the contribution due to the bipolaron, $\sigma_b$, and the one due to the ions closest to it, $\sigma_{\mathrm{loc}}$, isolating the cross-correlation contribution between the two, $\sigma_{\mathrm{cross}}$:
\begin{align}
    \overline{\sigma}(\lambda) = \sigma_b + \sigma_{\mathrm{loc}}(\lambda) + \sigma_{\mathrm{cross}}(\lambda), \\
    \sigma_b \propto \int_0^\infty \langle \mathbf{J}_b(t)) \cdot \mathbf{J}_b(0) \rangle \dd{t}, \\
    \sigma_{\mathrm{loc}}(\lambda) \propto \int_0^\infty \langle \mathbf{J}_{\mathrm{loc}}(\lambda, t) \cdot \mathbf{J}_{\mathrm{loc}}(\lambda,0) \rangle \dd{t}. \label{eq:sigma loc}
\end{align}
Excluding $\sigma_{\mathrm{cross}}$ entails neglecting cross-correlation contributions between the ions and the bipolaron.
The value of ${\sigma_b + \sigma_{\mathrm{loc}}}$ as a function of the ratio between $\lambda$ and the average Wannier spread of the bipolaron along the dynamics, $\varsigma_{\mathrm{avg}}$, is shown in Fig.~\ref{fig:cross conductivity}. When $\lambda/\varsigma_{\mathrm{avg}} \ll 1$, $\mathbf{J}_{\mathrm{loc}}$ is small because only few to no ions are within the cutoff distance from the bipolaron, and $\sigma_{\mathrm{cross}}$ approaches zero, thus $\overline{\sigma}$ and ${\sigma_b + \sigma_{\mathrm{loc}}}$ are indistinguishable. Around $\lambda=\varsigma_{\mathrm{avg}}$, the local charge flux includes the contributions due to the ions that are closest to the bipolaron, and nothing else. Therefore, the correlation among the local charge flux and the bipolaron's is maximal. When $\lambda$ is sufficiently large with respect to $\varsigma_{\mathrm{avg}}$ (i.e., $\lambda \gtrsim 1.5 \varsigma_{\mathrm{avg}}$), correlation effects tend to vanish, as the local charge flux becomes equivalent to the global one. In fact, $\overline{\sigma}$ and ${\sigma_b + \sigma_{\mathrm{loc}}}$ become again compatible within error bars.
The PBE0 results display a larger degree of correlation between ions and the bipolaron compared to the PBE ones: in fact, the relative difference between $\overline{\sigma}$ and ${\sigma_b + \sigma_{\mathrm{loc}}}$ becomes as large as $74\%$ for PBE0 at $\lambda=\varsigma_{\mathrm{avg}}$, while it stays at $19\%$ for PBE. Once again, this can be explained by the effect of EXX, that strengthens the interactions betwteen the ions and the bipolaron.

\section{Conclusions}\label{sec:conclusions}

In this work, we have explored the impact of self-interaction on the structural and transport properties of dissolved electrons in non-stoichiometric molten salts through MD simulations of a binary NaCl melt with excess Na. We have found that, when EXX is taken into account, the bipolaron exhibits a tendency to localize within well-defined solvation cells, whereas the RPDF appears structureless when using a semi-local functional. The localization  of the bipolaron in the solvation cell is manifested as a statistical property, rather than an instantaneous one, as the RPDFs computed separately on each step of the trajectory are uncorrelated with the bipolaron's spatial extent, both statically and dynamically. The distribution of the bipolaron's spread, entailing its spatial extension, testifies the larger average degree of localization in the PBE0 simulation, in accordance to the reduced self-interaction induced by the presence of EXX. 

The implications of these observations on the ionic transport properties of the melt are substantial. Notably, the inclusion of EXX significantly decreases the electrical conductivity. On a local level, charge transport due to ions alone correlates with the bipolaron's motion. This feat notwithstanding, this correlation dissipates on larger scales, resulting in a total electrical conductivity that can be decomposed into a purely ionic contribution, rationalized through integer and constant atomic OSs, and a purely bipolaronic contribution, associated with the motion of the WCs related to the HOMO.

Our study sheds light on the intricate interplay between charge and mass transport in non-stoichiometric molten salts, highlighting the importance of accurately accounting for self-interaction in simulations to capture the underlying mechanisms. By confirming the nontrivial regime where charge and mass transport are effectively uncorrelated, this work represents a first step towards further investigations into the fundamental processes governing the transport properties in complex fluids. 

\section*{Data Availability}
The data, sample input files, and data-analysis scripts that support the plots and relevant results within this paper are available on the Materials Cloud platform.\cite{talirz2020materials} See DOI:\url{https://doi.org/10.24435/materialscloud:8f-d7}.

\section*{Author Declarations}
The authors have no conflicts to disclose.

\section*{Acknowledgements}

We thank A. Grisafi, K. Rossi, and D. Tisi for fruitful discussions, and G. Fraux for technical support. This work was partially supported by the European Commission through the \textsc{MaX} Centre of Excellence for supercomputing applications (grant number 101093374) and by the Italian MUR, through the PRIN project \emph{FERMAT} (grant number 2017KFY7XF) and the Italian National Centre from HPC, Big Data, and Quantum Computing (grant number CN00000013). FG acknowledges funding from the European Union's Horizon 2020 research and innovation programme under the Marie Sk\l{}odowska-Curie Action IF-EF-ST, grant agreement number 101018557 (TRANQUIL).

\appendix

\section{Data-driven methods}\label{app:data driven}

\begin{figure}[htb]
    \centering   
    \includegraphics[width=\columnwidth]{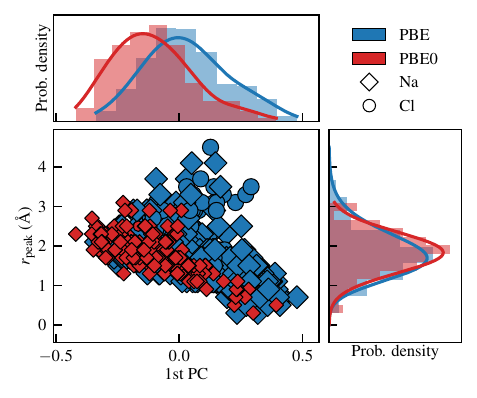}
    \caption{Correlation between the first PC of the bipolaron's local environments and the the distance of the first peak of the instantaneous Na-$b$ RPDF. Diamond-shaped markers represent configurations where the closest ion is a sodium, while circles represent configuration where the closest ion is a chlorine.}
    \label{fig:PCA distance}
\end{figure}

The distinction between the two functionals is further emphasized by a principal component analysis (PCA) conducted to investigate the presence of clusters within the features characterizing the local atomic environments of the system.\cite{goscinski2021role} To describe local environments we employed the widely used SOAP descriptors,\cite{bartok2013representing,de2016comparing,musil2021physics} with parameters provided in Tab.~\ref{tab:soap}, as implemented in \textsc{librascal}.\cite{musil2021efficient} Snapshots were sampled every $100\,\mathrm{fs}$ to ensure data uncorrelation. 
The SOAP descriptors for all the considered local environments and for both functionals were collected into a single feature matrix, $[\mathbf{X}]_{e,f}$, where rows represent local \textit{e}nvironments and columns represent SOAP \textit{f}eatures. The matrix, $\mathbf{X}$, was subsequently centered, and its covariance matrix, ${\mathbf{C} = \mathbf{X}^T \mathbf{X}}$, was diagonalized with eigenvalues sorted in descending order. The PCA was then obtained by projecting $\mathbf{X}$ onto the first $n\mathrm{PC}$ eigenvectors of $\mathbf{C}$, thereby accounting for as much variance in $\mathbf{X}$ as possible while simultaneously reducing its dimensionality.

Fig.~\ref{fig:PCA distance} illustrates the correlation between the first PC, which accounts for 17\% of the variance, and the distance of the first peak of the instantaneous Na-$b$ RPDF, $r_\mathrm{peak}$. The Pearson correlation coefficients between these quantities for PBE and PBE0 are $-0.43$ and $-0.69$, respectively. Notably, the correlation for the PBE data gets to $-0.55$ when considering only structures where Na is the closest ion. This suggests that the first PC of the bipolaron's SOAP descriptors is related to the ionic species and the proximity of its closest neighbor. The correlation is more robust for the PBE0 calculation, further corroborating the fact that the inclusion of EXX influences the first solvation shell of the bipolaron.

\begin{table}[htb]
    \centering
    \begin{tabular}{ll}
        \toprule
        Keyword & Value \\
        \midrule
        \texttt{`soap\_type'} & \texttt{`PowerSpectrum'} \\
        \texttt{`interaction\_cutoff'} & $6.5$ \\
        \texttt{`max\_radial'} & $12$ \\
        \texttt{`max\_angular'} & $10$ \\
        \texttt{`gaussian\_sigma\_constant'} & $0.3$ \\
        \texttt{`gaussian\_sigma\_type'} & \texttt{`Constant'} \\
        \texttt{`cutoff\_smooth\_width'} & $0.5$ \\
        \texttt{`radial\_basis'} & \texttt{`GTO'} \\
        \texttt{`inversion\_symmetry'} & \texttt{True} \\
        \texttt{`normalize'} & \texttt{True} \\
        \bottomrule
    \end{tabular}
    \caption{SOAP hyper-parameters used to conduct the PCA of the bipolaron's local environments.}
    \label{tab:soap}
\end{table}

\section{Computational details}\label{app:computational}

MD simulations are performed with the \textsc{cp2k} code,\cite{hutter2014cp2k} version 9.1. The PBE and PBE0 functionals are parametrized according to the revised formulation also known as revPBE.\cite{zhang1998comment} The electronic density is expanded in the TZV2P-GTH Gaussian basis set. We employed G\"odecker-Teter-Hutter (GTH) pseudopotentials\cite{goedecker1996separable} encompassing electrons lying in the second shell, allowing for polarization effects which may contribute significantly to the accuracy of the simulations.\cite{ishii2015transport} Gaussian functions are mapped to a multi-grid with four levels; the plane-wave cutoff for the finest level is set to $400\,\mathrm{Ry}$, with a relative cutoff of $60\,\mathrm{Ry}$. Computations are sped up with the Auxiliary Density Matrix Method (ADMM).\cite{guidon2010auxiliary} The hybrid calculations employ a Coulomb operator truncated\cite{spencer2008efficient} at $6\,\text{\AA}$ to further reduce the computational effort while retaining its accuracy. Long-range Van der Waals interactions are modeled through Grimme's D3 corrections.\cite{grimme2010consistent} All calculations are spin-polarized in the singlet state, which is energetically favored.~\cite{pegolo2020oxidation, kristoffersen2018chemistry}

Both MD simulations sample the canonical ensemble at $1300\,\mathrm{K}$ through the Bussi-Donadio-Parrinello thermostat\cite{bussi2007canonical} with a time constant of $1\,\mathrm{ps}$, i.e. two thousands times the integration time-step of $0.5\,\mathrm{fs}$. The Kohn-Sham wavefunctions are transformed to the basis of MLWFs~\cite{marzari2012maximally} through Jacobi rotations every $1\,\mathrm{fs}$ to collect WCs and spreads to be used to compute the charge flux of Eq.~\eqref{eq:J wannier}.
The PBE simulation has been thermalized for $5\,\mathrm{ps}$. The production run is $50\,\mathrm{ps}$ long. The PBE0 simulation has been initialized from a snapshot drawn from the equilibrated PBE simulation, and then further thermalized for $2\,\mathrm{ps}$. The production run is $15\,\mathrm{ps}$ long. 
Input files can be found in Materials Cloud.\cite{talirz2020materials}

\bibliography{main}

\end{document}